\documentclass[%
letterpaper,
reprint,
superscriptaddress,
showkeys,
showpacs, preprintnumbers,
 amsmath,amssymb,
 aps,
 prl, twocolumn,
]{revtex4-1}

\usepackage{graphicx}
\usepackage{dcolumn}
\usepackage{bm}
\usepackage[mathlines]{lineno}

\begin{document}

\title{Experimental study of the microwave emission from electrons in air}

\author{E.~Conti}
 \email{Corresponding author: enrico.conti@pd.infn.it}
\affiliation{ I.N.F.N., sez. di Padova, Via F. Marzolo 8, Padova, Italy}
\author{G.~Collazuol}
\affiliation{Dip. di Fisica e Astronomia ``G.~Galilei'', Univ. di Padova,Via F.~Marzolo 8, Padova, Italy}
\affiliation{ I.N.F.N., sez. di Padova, Via F. Marzolo 8, Padova, Italy}
\author{G.~Sartori}
\affiliation{Dip. di Fisica e Astronomia ``G.~Galilei'', Univ. di Padova,Via F.~Marzolo 8, Padova, Italy}
\affiliation{ I.N.F.N., sez. di Padova, Via F. Marzolo 8, Padova, Italy}

\date{\today}

\begin{abstract}
We searched for the emission of microwave radiation in the Ku band generated by a 95 keV electron beam in air. We unequivocally detected the radiation, and measured its yield and angular dependence.
Both the emitted power and its angular pattern are well described by a model, where microwave photons are generated via bremsstrahlung in the free-electron atomic-nucleus collisions,  during the slowdown of the electrons. As a consequence, the radiation is not isotropic but peaked in the forward direction.

The emission yield scales proportionally  with the  number of electrons. This contrasts a previous claim that the yield scales with the number squared, due to coherence.

With a Monte Carlo simulation we extrapolate our results to the Ultra High Energy Cosmic Ray energy range. 

\end{abstract}



\maketitle

The nature of  Ultra High Energy Cosmic Rays (UHECR) (energy $> 10^{18}$~eV) is still one of the most intriguing mysteries of the universe. 
The experimental study of this topic
is challenging, since the rate of UHECRs is extremely low, $\lesssim 1$~event/$100km^2$/year. Very large areas must be instrumented to detect only a few events. Experiments such as the Pierre Auger Observatory  \cite{AUGER} in the southern hemisphere, and Telescope Array \cite{TA} in the northern one, have extensions of the order of $1000~\text{km}^2$,  and are nowadays the only  observatories dedicated to this subject. 
The technology exploited so far has almost reached   its limit  and  can hardly be scaled up to increase  the experimental site extension by one more order of magnitude. New techniques must be envisaged, such, for example, those based on the near infrared fluorescence \cite{conti} or on the microwave detection.

The emission of microwave (MW) radiation by an EAS (Extended Air Shower) traveling in air is due to three mechanisms: Cherenkov effect (both incoherent and coherent), synchrotron effect (geomagnetic or geosynchrotron effect), and bremsstrahlung. While the first two  are well established on the theoretical and the experimental ground (see for example refs.\cite{Jelley1958, Askaryan, SLAC2001,ANITA2007} for Cherenkov, and refs.\cite{KahnLerche,Falcke2005, CODALEMA2009,Huege} for the geomagnetic effect), the emission via bremsstrahlung has not been investigated accurately with experimental measurements.

When free electrons knock on the atomic nuclei and electrons of air, 
they emit photons via bremsstrahlung over a wide wavelength range, including radio-frequency (RF) and MW. The first studies were performed by Bekefi et al. \cite{Bekefi1959, Bekefi1966}, who observed the radiation emitted by the plasma in a glow discharge lamp inside a RF cavity. In 1969,   the yield of bremsstrahlung radiation from EAS was estimated \cite{Jelley1969} to be from 10 to 100 times higher than the incoherent Cherenkov effect, but still too low, by several orders of magnitude, to be detectable. The bremsstrahlung radiation in air has not been investigated till the recent research of Gorham et al. \cite{Gorham}.
They claimed to have detected the MW signal from electromagnetic showers, generated by a 28~GeV  electron beam, in the 1.5$\div$6~GHz frequency range, and to have evidence of a coherent mechanism, which enhances the yield. The proof of the coherence was based on the measured quadratic dependence of the yield upon the beam energy (or, equivalently, the number of shower particles). 
 Their major issue was the incoherent and fully polarized Cherenkov radiation, which was about 3 orders of magnitude higher than the searched signal.
 The explanation of the results follows the model introduced in ref.\cite{Bekefi1966}, where the ionized air  is treated as an uniform plasma with a low degree of ionization at thermal equilibrium with isotropic,  maxwellian, and steady-state velocity distribution. The resulting emission is isotropic.

These results, scaled to actual EAS, imply that UHECRs can be detected with cheap radio-frequency techniques, and stimulated many experiments aiming at the MW detection of high energy electrons \cite{AMY} and UHECRs \cite{Gorham,MIDAS,CROME}. After long periods of data taking without evidence of signals correlated with EAS,  many doubts arose, not only about the power flux, but also about the existence itself of the  emission \cite{MIDAS2}.

In this article we present new measurements of the MW emission from a 95 keV electron beam in air in the Ku band ($\sim 11$GHz). The advantages of using a low energy beam are: i) the electrons are below the Cherenkov threshold, both in air and in the materials crossed by the beam, so that the main source of background radiation is definitively avoided;  ii) the range of such electrons in air is limited to about 10~cm, so the setup is compact and the experimental conditions are easily kept under control. 
The main drawback consists in a rather weak signal intensity. To our knowledge, these are the first measurements of the MW bremsstrahlung in a background free, controlled and repeatable environment.

\begin{figure}
\centering
\includegraphics[width=8.6cm]{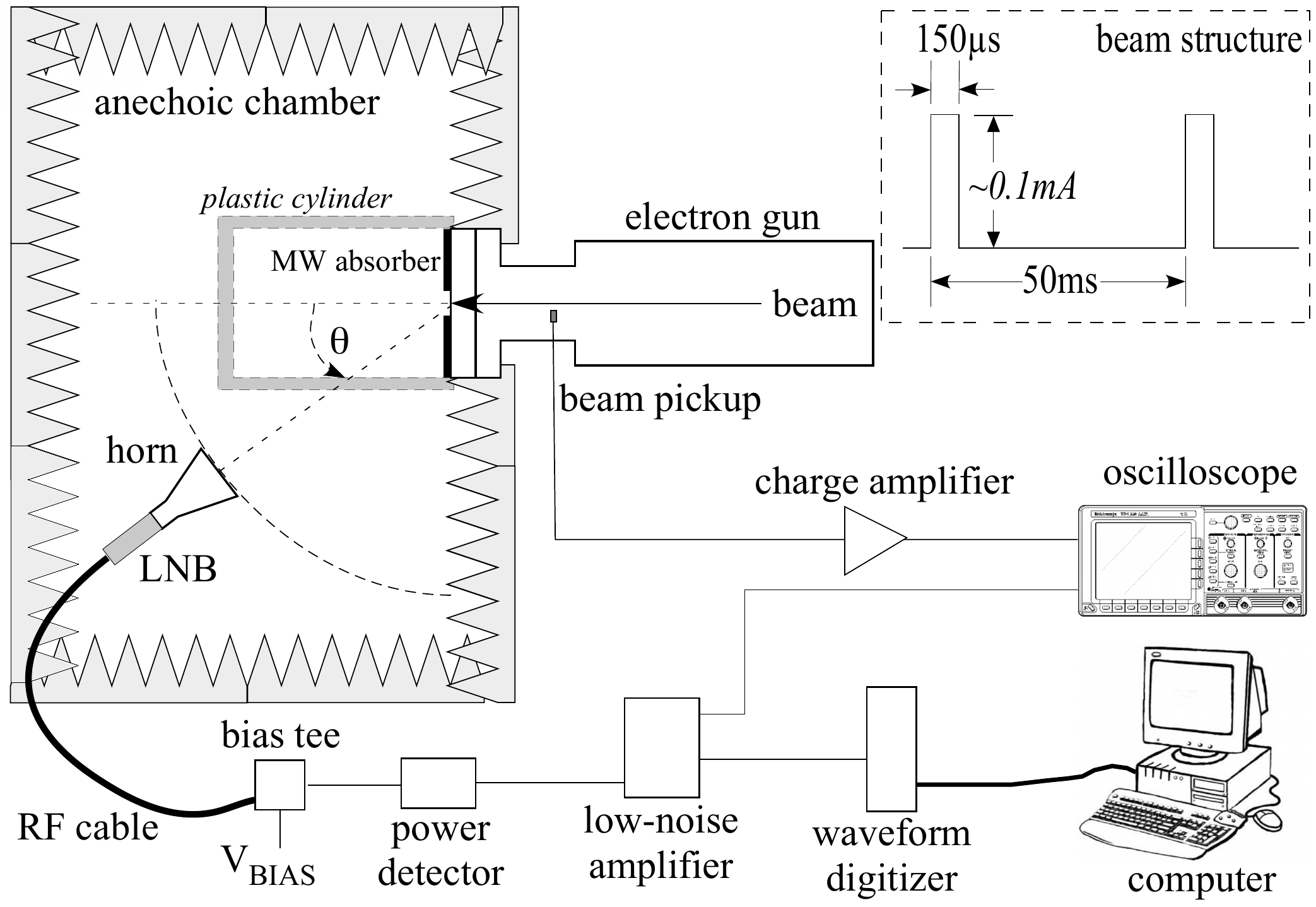}
\caption{A schematic view of the experimental setup (not to scale). The cylindrical plastic  vessel is also shown, which is evacuated and used to measure possible spurious signals, and then removed during all other measurements.}
\label{fg:setup}
\end{figure}

A schematic of the experimental setup is shown in fig.\ref{fg:setup}.
An electrostatic gun (Kimball Physics Inc.) accelerates electrons
up to $95$ keV kinetic energy, in pulses $150 \mu s$ long. Electrons exit the gun in air through a synthetic diamond window, 
$20\mu m$ thick. In air, the electron energy spectrum has the typical Landau distribution, with most probable energy of 81 keV. The beam is monitored by measuring each pulse intensity with a beam pick-up, internal to the electron gun. The  relationship between the pick-up signal and the current in air after the diamond window is measured by intercepting and collecting the whole beam with a metallic stopper.

The MW signal is detected by a Low Noise Block (LNB) (Norsat mod. 1000H) 
operating from 10.95 to 11.70 GHz, with an internal 10 GHz Local Oscillator, and  nominal gain of 60 dB.  The LNB feed is a pyramidal horn with 20 dB nominal gain, positioned  inside an anechoic chamber made by pyramidal RF absorbers. The horn points to the center of the exit flange, at a constant distance from it, forming an angle $\theta$ with respect to the beam direction. The LNB has a single dipole antenna and detects one polarization each time.
Given the horizontal plane individuated by the beam axis, we choose either the polarization perpendicular to that plane, or the one lying in the plane.
The exit flange on the electron gun is covered with a thin RF absorber foil to avoid reflections of the MW radiations. The MW signal, triggered by the beam pulse, is integrated by a power detector (Mini-Circuits mod. ZX47-60LN), then amplified and shaped by a low-noise amplifier (Stanford Research System mod. SR560), before being acquired by a 14-bit, 100-MHz-clock waveform digitizer (CAEN mod.1728A), and stored on disk for the offline analysis. A good signal-to-noise ratio requires to average at least  $10^4$ pulses. In such conditions, the power sensitivity is of the order of $10^{-16}$~W.

The power detector voltage output  $V_{PD}$ has a logarithmic dependence on the MW input power $P_{MW}$: $V_{PD} = a + b\cdot \ln(P_{MW})$. The microwave signal is the sum of a constant power level $P_{BB}$, characteristic of the blackbody radiation at room temperature, and of a much smaller level $P_{\Delta}$, correlated to the beam pulse: $P_{MW} = P_{BB} + P_{\Delta}$. Since $P_{\Delta} \ll  P_{BB}$ we can express the power detector output as $V_{PD} = V_{BB} + V_{\Delta}$, with $V_{BB}=a + b\cdot \ln(P_{BB})$ and $V_{\Delta}= b\cdot {P_{\Delta}/P_{BB}} = {b\cdot P_{\Delta}\exp[-(V_{BB}-a)/ b]}$. 
The electronic chain, from the RF cable to the waveform digitizer, is calibrated injecting a monochromatic signal, with known power and frequency, and with the same width of the beam pulse, generated by a calibrated signal generator (Rohde $\&$ Schwarz mod. SMW200A). The acquired waveforms are then analyzed offline with the same procedure as for real signals. With our setup, 100 ADC counts correspond to a power level, at the input of the power detector, of about 1 nW above the blackbody noise level, which is 6 orders of magnitude higher.

The LNB-horn system is calibrated by exposing it to the pyramidal RF absorbers, which are an almost ideal blackbody source at room temperature with unit emissivity, and accounting for the angular response of the horn.
The result is consistent with the LNB nominal gain and a horn aperture efficiency $\simeq 0.6$.

The horn solid angle is determined by the position of the horn phase center, 
which is established experimentally  measuring the variation of the signal $V_{\Delta}$ generated by the pulsed beam as a function of the distance $D$, and fitting the data with the law $V_{\Delta}(D) = V_{\Delta0}/(D-D_0)^2$. The position is known with $\pm1$ cm error.

Two approaches were pursued to verify that the signal detected by the LNB is not spurious, but generated unambiguously by the electrons in air. First,  a thin 
cardboard, transparent to the MW radiation, was placed in contact with the diamond window, to dump the electron avoiding their propagation in air.
Alternatively,  a plastic cylindrical vessel was added at the end of the electron gun (see fig.\ref{fg:setup}), and evacuated. The electrons, after the diamond window, propagated in vacuum and were stopped by the vessel bottom. In both cases, no MW emission was observed.

\begin{figure}[!t]
\centering
\includegraphics[width=8.6cm]{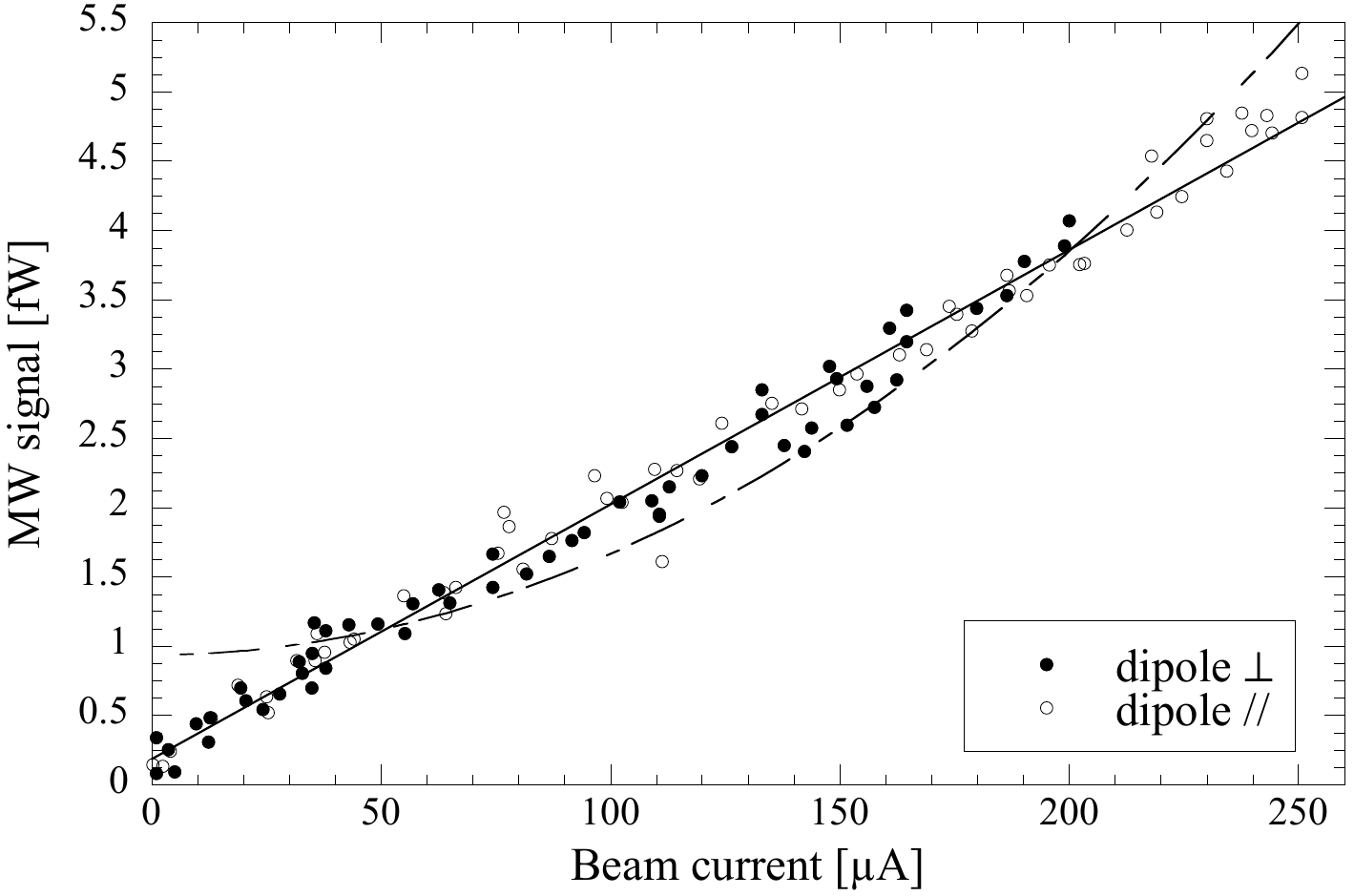}
\caption{Dependence of the microwave signal from the beam intensity, at $\theta=25^\circ$, for the two   orientations of the LNB dipole. The best least squares fits of all data are shown with a linear law (continuous line) and with a quadratic law  $f(x)=A+B\cdot x^2$  (dashed line). }
\label{fg:linearity}
\end{figure}

The dependence was measured of the MW signal upon the beam current, at a fixed angle $\theta$, for both LNB polarizations. The result is shown in fig.\ref{fg:linearity} at $\theta=25^\circ$, but similar plots are obtained at different angles. The data of the two polarizations, which spans two orders of magnitude in the beam current, are compatible within the errors, proving that the radiation is not polarized. The least squares regressions with a linear and with a quadratic law ($f(x)= A + B\cdot x^2$, analogous to \cite{Gorham}) are also plotted. While the linear law interpolates well the data, the quadratic law fits poorly. 
We calculated the a posteriori error as $\sigma^2 = \sum_{i=1}^N [y_i-f(x_i)]^2/(N-2)$  for the two regressions, $\sigma_1$ (for the linear fit) and $\sigma_2$ (for the quadratic fit), and their errors $\sigma_{\sigma_1}=\sigma_1/\sqrt{2N}$ and $\sigma_{\sigma_2}=\sigma_2/\sqrt{2N}$, where $N$ is the number of data. 
The ratio  
$
\Delta = {{|\sigma_1 - \sigma_2|} / {\sqrt{\sigma_{\sigma_1}^2+\sigma_{\sigma_2}^2} }}
$ determines the compatibility of the a posteriori errors.
We find $\sigma_1 \approx 0.5\cdot\sigma_2$ and $\Delta = 7.0$  which rule out any quadratic dependence of the microwave yield on the beam intensity.

\begin{figure}[!b]
\centering
\includegraphics[width=8.6cm]{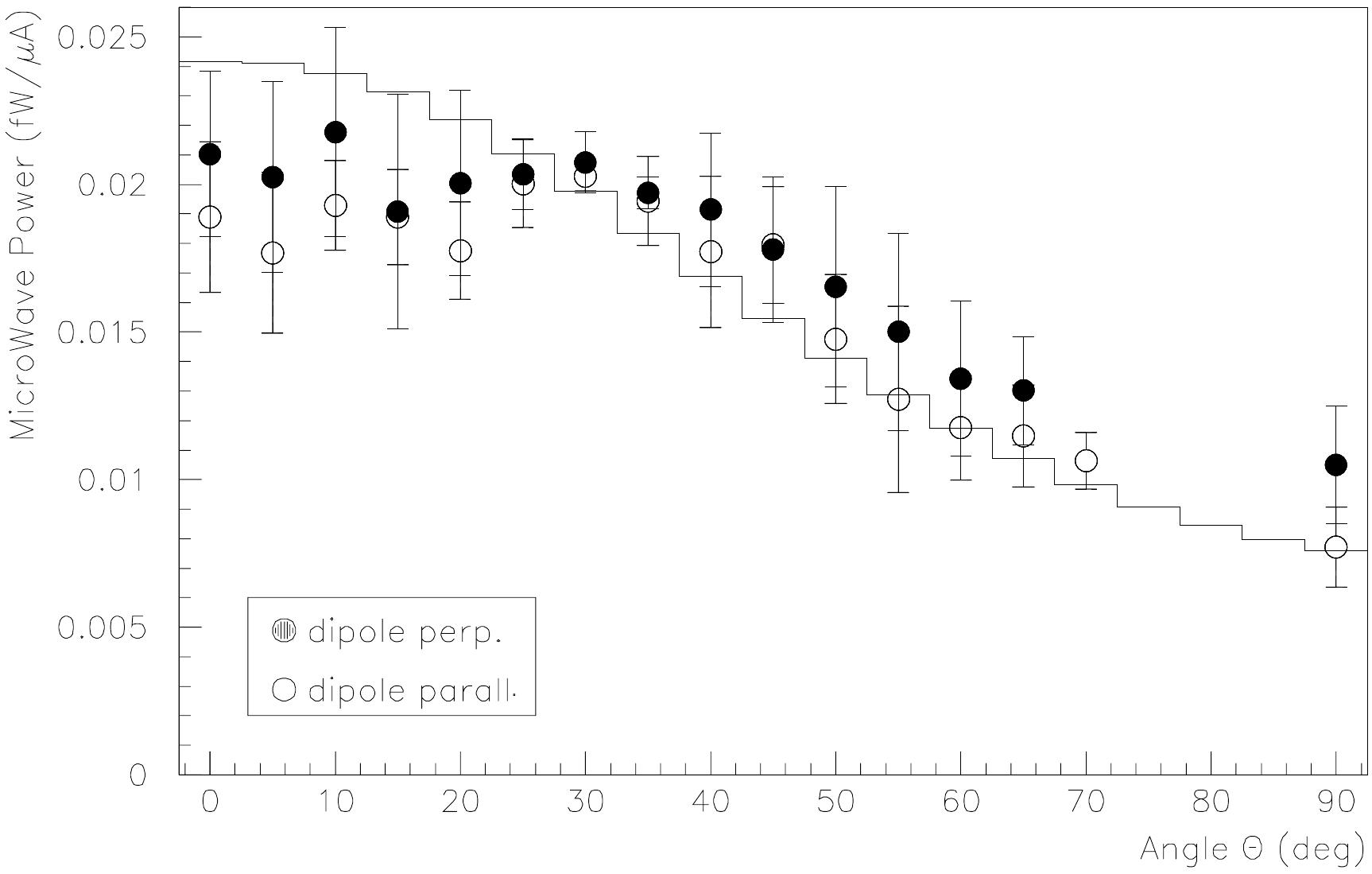}
\caption{Angular dependence of the  MW emitted power for the two orientations of the LNB dipole, and result of the Monte Carlo simulation (full line). The error bars represent the measurement errors. Data from $70^\circ$ to $85^\circ$ could not be taken because the horn hit against  the wall of the anechoic chamber.}
\label{fg:angular}
\end{figure}

The angular dependence of the MW yield gives insights on the physical process responsible of the emission. Our measurements are shown in fig.\ref{fg:angular} for both dipole orientations, which agree within the experimental errors, demonstrating once more that the radiation is  unpolarized.  The emission depends on the observation angle $\theta$, with a difference of about a factor 2 between $\theta=0^\circ$ and $\theta=90^\circ$. Such dependence 
is  ascribable mainly to the fact that
the MW radiation is not generated isotropically along the electron tracks, having instead characteristic angular distributions, as discussed in the following.  

In order to follow in detail the shower evolution and the radiation emission on a step-by-step basis,  a Monte Carlo code was written, based on the PENELOPE package \cite{penelope}, which is adequate to describe electromagnetic interactions down to an energy of 100~eV.
 Differently from ref. \cite{Gorham}, we track the electrons from their production to thermalization, without assuming any hypothesis on equilibrium distributions and on time-stationarity.

Although  bremsstrahlung processes are considered to be well understood, quantitative calculations of cross sections and angular distributions require approximations which depend on the electron initial kinetic energy and the emitted photon energy. A compilation of such calculations can be found in ref. \cite{koch} and in the more recent reviews \cite{nakel, haug}. Tabulated data of the ``scaled" cross section $\sigma_{\text{scaled}}$:
\begin{eqnarray}
    \label{eq:sigmascaled}
\sigma_{\text{scaled}} = \frac{\beta^2} {Z^2}~k~\frac {d\sigma} {dk}(Z,T,k)
\end{eqnarray}

\noindent
are found in refs.\cite{Pratt,Seltzer}, for electron kinetic energy $T$ from 1 keV to 10 GeV, where $\beta = v/c$ is ratio between the electron and light velocity,
 $Z$ is the nucleus atomic number, $k$ the photon energy, and $d\sigma/dk$ the bremsstrahlung differential cross section.

The total cross section $\sigma$ is obtained by integrating eq.~(\ref{eq:sigmascaled}): $\sigma(T) = (Z^2/\beta^2) (\Delta \nu/\nu_0)\sigma_{\text{scaled}}$, where $\nu_0$ and $\Delta \nu$ are the central frequency and the bandwidth of the LNB, respectively, and ${\Delta \nu/\nu_0}\ll1$. For air, in the 1$\div$100~keV range, $\sigma$ scales approximately as $1/T^{0.78}$. In the region $T$=1$\div$10~keV  $\sigma$ is of the order of few barns. The corresponding  mean free path $\lambda = 1/n\sigma \approx 10^4$cm ($n$ is the air number density) implies that the probability $p$ of a MW emission along the electron path $L$
 is low: $p = L/\lambda \approx 10^{-4}$. The expected MW power $P_{MW}$ irradiated by a current $I$ is therefore $P_{MW} = p\cdot(I/q) \cdot h\nu_0 \cdot \Omega_{\text{horn}} \approx $ fW, for $I=100\mu A$
and a horn geometrical acceptance $ \Omega_{\text{horn}}  \approx 1\%$ ($q$ is the electron charge, $h$ the Planck constant).

The bremsstrahlung angular distribution $S(Z,T,k,\alpha)=\frac{d\sigma}{dkd\Omega}/\frac{d\sigma}{dk}$ is calculated in ref. \cite{Tseng1979} for the 1$\div$500 keV energy range and parametrized as:
$S(Z,T,k,\alpha)= (A/4\pi) \sum_{i=0}^5 B_iP_i(\alpha)/(1-\beta \cos\alpha)^4$,
where the coefficients $B_i(Z,T,k)$ are tabulated, $P_i(\alpha)$ is the i-th Legendre polynomial, $\alpha$ is the angle between the electron and photon momenta, and $A$ is defined by the normalization condition $\int Sd\Omega =1$.
For  $T<1$ keV and  $k\ll T$, the angular distribution $S$ is flat over $\alpha$
due to the  atomic screening of the nucleus. At higher energies, $S$ 
tends to be peaked toward $\alpha =0$, because of  the dipole nature of the bremsstrahlung radiation.
Depending on the electron energy, multiple scattering  can smooth such directionality.

PENELOPE  follows all particles of the shower generated by the primary electron dividing their path in elementary steps of length $\ell_i$ where the electron energy is $E_i$. We calculate a quantity proportional to the MW emission yield $M = \sum\nolimits_{i} \ell_i \sigma(E_i) S(E_i,\alpha_i)\delta\Omega_i$, where  $\delta\Omega_i$ is the effective solid angle of the horn including its angular response, and the summation is done over the entire path of all charged particles. Then, the MW power  at the angle $\theta$ is
$
{P_{MW}(\theta) / I } = n M (h\nu_0 / q).
$
The electrons are simulated down to 1 keV, since the residual energy contribuites by less than $1 \%$ to the total radiated power, because the cross section $\sigma$ decreases rapidly \cite{kimpratt}.

The result of the simulation is plotted in fig.\ref{fg:angular} together with the experimental data.  
The experimental ($y_{exp}$) and simulated ($y_{MC}$) distributions are fitted to achieve the best agreement leaving  the normalization factor $\kappa$ as free parameter: $y_{MC}=\kappa \cdot y_{exp}$.  The minimum $\chi^2$ value is 56.4 for 30 degrees of freedom. The normalization factor is $\kappa=0.73\pm0.01$.
The error on $\kappa$ introduced by the calibration procedure amounts to $19.7\%$, due to the following contributions summed in quadrature: 13.2$\%$ from the beam current calibration, 11.8$\%$ from the phase center position, $7.0\%$ from the electronic chain calibration, 5.0$\%$ from the LNB-horn system absolute calibration,  $0.9\%$ from the waveform digitizer linearity.
The uncertainty on 
$\sigma_{\text{scaled}}$ is about $10\%$ \cite{Pratt,Seltzer} and dominates over other errors. Thus one can conclude that the model agrees well with the experimental measurements, both in yield and angular distribution.

Using the Monte Carlo simulation, the results can  be extended to the high energy range up to 10~GeV using the cross section (\ref{eq:sigmascaled}). The bremsstrahlung angular distribution $S$ was obtained analytically in ref. \cite{kohn} for $T\ge$ 1~keV integrating the Bethe-Heitler triple differential cross section \cite{BH}. However, a  simpler equation can be used when $k\ll T$:
\begin{eqnarray}
    \label{eq:esse2}
S(T,\alpha) = \frac{3}{16\pi} \frac{1-\beta^2} {(1-\beta \cos\alpha)^2} \Big[1+\Big(\frac{\beta - \cos\alpha} {1-\beta\cos\alpha}\Big)^2 \Big]
\end{eqnarray}

\noindent
which satisfies the normalization condition $\int Sd\Omega =1$. The expression was derived in the classical case \cite{jackson} and  agrees well with the exact formula for $k/T \approx 0$
\cite{kohn}. We point out that eq.~(\ref{eq:esse2}) foresees an emission strongly peaked in  forward direction for high energy electrons. 

\begin{figure}
\centering
\includegraphics[width=8.6cm]{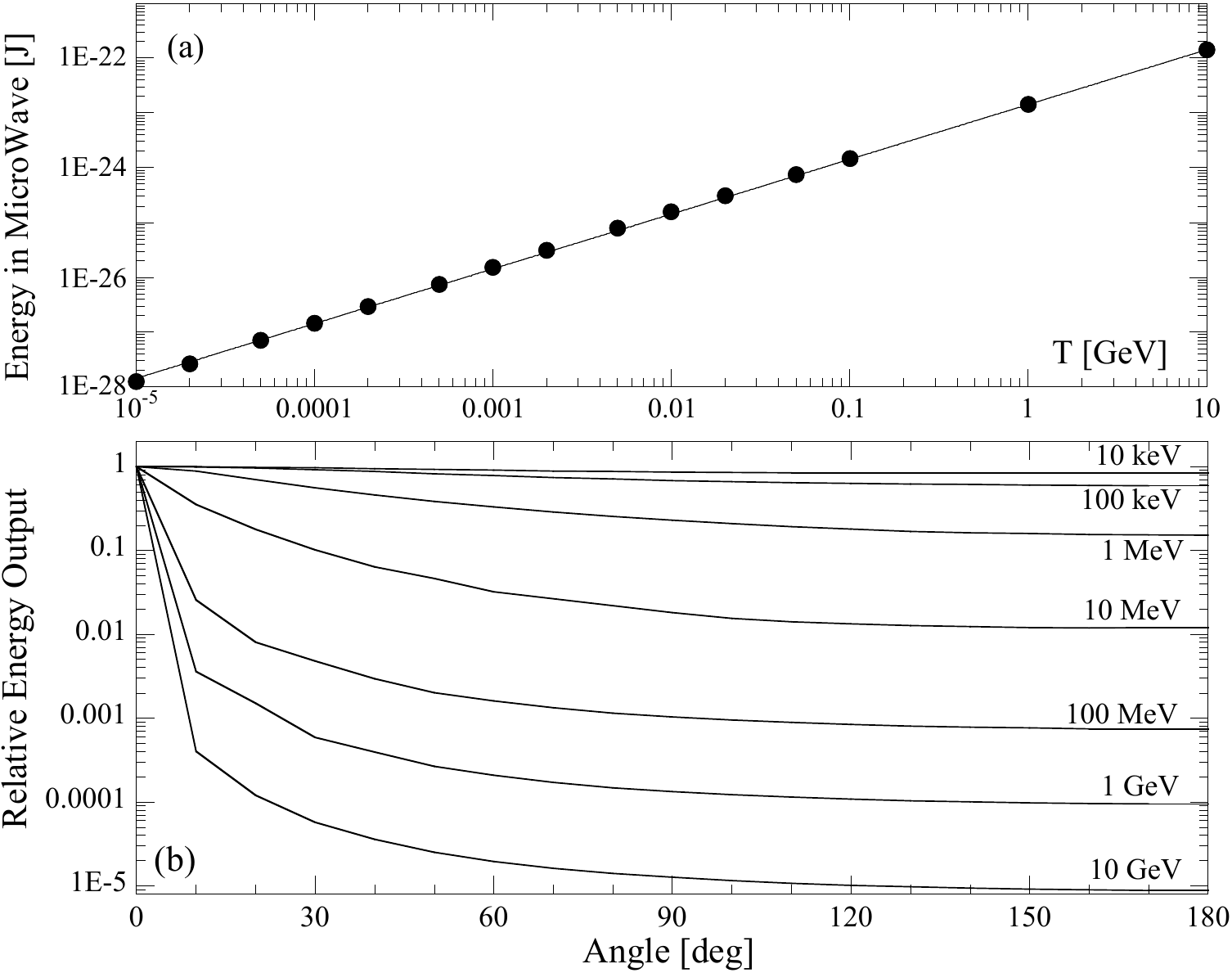}
\caption{Results of the Monte Carlo simulation.
(a). Dependency of the energy $\mathcal{E}_{MW}$ emitted in MW radiation on the electron energy $T$. The line is the best fit with a straight line $\mathcal{E}_{MW} = a\cdot T$, with $a = 1.420\cdot10^{-23}$J/GeV.
(b). Angular shape of the electromagnetic shower at different energies.  
}
\label{fg:EMW}
\end{figure}

Since the detectable energy of a hadronic shower in air is determined entirely by its electromagnetic component \cite{matthews},  hadronic EAS can be treated as composed by electromagnetic showers.
The result of the simulation for pure electromagnetic showers is shown in fig.\ref{fg:EMW}(a), where 
the dependence of the total energy emitted as microwave radiation $\mathcal{E}_{MW}$ in our frequency range is plotted as a function of the initial electron energy $T$. The fraction of energy transferred to MW in our wavelength region is very small, about $0.9\cdot10^{-13}$. The  angular pattern strongly favors the emission in the forward direction (fig.\ref{fg:EMW}(b)). At 1~GeV, for instance, the emission is almost entirely contained within an angle of $10^\circ$ with respect to the initial trajectory. 
Although qualitative for hadronic showers, the simulation indicates that the emission is not isotropic but has a preferred forward direction.

The minimum noise level required by the MW detection scheme is found as follows. Signals are detected with a signal-to-noise ratio SNR if the minimum detectable flux rate of the apparatus $\Delta\Phi_{min}$ is at least 
$
\Delta\Phi_{min} = (f~\mathcal{E}_{MW}/\Delta t)/(\text{SNR}~A_e \Delta\nu)
$
where $A_e$ is the antenna effective aperture, $f$ the fraction of energy which arrives to the antenna,  and $\Delta t$ the duration of the signal. $\Delta\Phi_{min}$ can be  expressed in term of the total system noise temperature $\mathbb{T}$ as 
$
\Delta\Phi_{min} = (2K_S/A_e)k_B\mathbb{T}/ \sqrt{\Delta\nu~\tau}
$
 \cite{kraus}, where $\tau$ is the integration time, in this case equal to $\Delta t$, $k_B$ the Boltzmann's constant, and $K_S$ the sensitivity constant, which depends on the type of receiver, and it is equal to 1 for total-power receivers.  Therefore, the required noise temperature  is
 $
 \mathbb{T} = (1/2k_B\sqrt{\Delta t\Delta\nu})(f\mathcal{E}_{MW}/\text{SNR})
 $.

The factor $f$ depends on the particular experimental setup. Consider an antenna with a radius of 10~m and an angular acceptance of $20^\circ$ (solid angle $\approx 1\%$), which is larger than those used in refs.\cite{Gorham,MIDAS,MIDAS2}.
For a UHECR with $T = 5\cdot10^{18}$~eV, $\mathcal{E}_{MW}=7\cdot10^{-14}$~J. If we look at the shower from the side, the transit time is $\Delta t \approx 100~\mu s$. $f$ is  the product of  the antenna solid angle and of the fraction of MW emitted at $90^\circ$, which is $\lesssim10^{-5}$, as appears from fig.\ref{fg:EMW}(b). Hence, a signal-to-noise ratio SNR~=~10, for example, requires $\mathbb{T} \lesssim$~1~K. This is probably the reason why the MW detection of EAS has been unsuccessful.

Should  the antenna point to an UHECR source, the EAS would be viewed from the front. Therefore $\Delta t~\approx~10~ns$, and the factor $f$ is determined mainly by the ratio between the antenna and the front shower areas. Conservatively, for a front shower radius  of the order of km, $f  \approx 10^{-6}$.  In the same conditions as above, $\mathbb{T} \gtrsim 100$~K, which is easily achievable. 

\begin{acknowledgments}
The authors acknowledge the skillful help of G.~Viola (Dip. di Fisica e Astronomia ``G.~Galilei'', Univ. di Padova).
\end{acknowledgments}


\begin{thebibliography}{00}

\bibitem{AUGER}
J. Abraham et al. (The Pierre Auger Collaboration), Nucl. Instr. and Meth. in Phys. Res. A 523 (2004), 50

\bibitem{TA}
T. Abu-Zayyad et al. (The Telescope Array Collaboration), Nucl. Instrum. and Meth. in Phys. Res. A 689 (2012), 87; H. Tokuno et al. (The Telescope Array Collaboration), Nucl. Instrum. and Meth. in Phys. Res. A 676 (2012), 54

\bibitem{conti}
E. Conti, G. Sartori, G. Viola, Astrop. Phys. 34 (2011), 333

\bibitem{Jelley1958}
J.V. Jelley, Suppl. Nuovo Cim., serie X, vol.8 (1958), 578

\bibitem{Askaryan}
G. Askaryan, Sov. Phys. JETP 14 (1962), 441
; 21 (1965), 658

\bibitem{SLAC2001}
D. Saltzberg et al., Phys. Rev. Lett. 86 (2001),  2802

\bibitem{ANITA2007}
P. W. Gorham et al. (ANITA Collaboration), Phys. Rev. Lett. 99 (2007), 171101

\bibitem{KahnLerche}
F.D. Kahn, I. Lerche, Proc. Roy. Soc. A 289 (1966), 206

\bibitem{Falcke2005}
H. Falcke et al., Nature 435 (2005), 313

\bibitem{CODALEMA2009}
D. Ardouin et al., Astrop. Phys. 31 (2009), 192

\bibitem{Huege}
T. Huege, Nucl. Instrum. and Meth. in Phys. Res. A, 604 (2009), S57

\bibitem{Bekefi1959}
G. Bekefi, J.L. Hirshfield, Sanborn C. Brown, Phys. Rev. 116 (1959), 1051

\bibitem{Bekefi1966}
G. Bekefi, \textit{Radiation Processes in Plasmas}, Wiley, New York, 1966

\bibitem{Jelley1969}
W.N. Charman, J.V. Jelley, Nuovo Cim. B, 63 (1969), 473

\bibitem{Gorham}
P.W. Gorham et al., Phys. Rev. D 78, 032007 (2008)

\bibitem{AMY}
J. Alvarez-Mu\~{n}iz et al., Nuovo Cim. C 36 (2013), 134

\bibitem{MIDAS}
J. Alvarez-Mu\~{n}iz et al.  Nucl. Instrum. and Meth. in Phys. Res. A 719 (2013), 70

\bibitem{CROME}
R. Smida et al., Proceedings of the 32nd International Cosmic Ray Conference, Beijing, China, 2011; arXiv:1108.0588

\bibitem{MIDAS2}
J. Alvarez-Mu\~{n}iz et al., Phys. Rev. D 86, 051104(R) (2012)
  
 
 \bibitem{penelope}
 www.oecd-nea.org/tools/abstract/detail/nea-1525; F. Salvat et al.,  Proceedings of a Workshop/Training Course,  OECD/NEA 5-7 November 2001
  NEA/NSC/DOC (2001) 19; 
 J. Sempau et al., Nucl. Instrum. and Meth. B 132 (1997) 377; 
J. Sempau et al.,  Nucl. Instrum. and Meth. B 207 (2003) 107

\bibitem{koch}
H.W. Kock, J.W. Motz, Rev. Mod. Phys. 31 (1959), 920

\bibitem{nakel} 
W. Nakel, Phys. Rep. 243 (1994), 317

\bibitem{haug} E. Haug, W. Nakel, \textit{The elementary process of bremsstrahlung}, World Scientific Publishing, 2004

\bibitem{Pratt}
R.H. Pratt, H.K. Tseng, C.M.Lee, Lynn Kissel, Atom. Data and Nucl. Data Tab. 20 (1977), 175

\bibitem{Seltzer}
S.M. Seltzer, M.J. Berger, Nucl. Instrum. and Meth. in  Phys. Res. B 12 (1985), 95

\bibitem{Tseng1979}
H.K. Tseng, R.H. Pratt, C.M. Lee, Phys. Rev. A 19 (1979), 187

\bibitem{kimpratt}
A. Florescu, O. I. Obolensky,  R.H.Pratt, J. Phys. B: At. Mol. Opt. Phys. 35 (2002), 2911.

\bibitem{kohn}
C. K\"{o}hn, U. Ebert, Atmosph. Res. 135-136 (2014), 432

\bibitem{BH}
H.A. Bethe, W. Heitler, Proc. Phys. Soc. Lond. 146 (1934), 83

\bibitem{jackson}
J.D. Jackson, \textit{Classical electrodynamics}, third edition, John Wiley \& Sons,  1998

\bibitem{matthews}
J. Matthews, Astrop. Phys. 22 (2005), 387

\bibitem{kraus}
J.D. Kraus, \textit{Radio Astronomy}, second edition, Cygnus-Quasar Books, 1986.

\end{thebibliography}
\end{document}